\newcommand{\rvac}{\rho_{\rm vac }}

\newcommand{\xib}{\mbox{\boldmath$\xi$}}

\newcommand{\dd}{{\rm d}}

\newcommand{\rl}{r_{\Lambda}}

\newcommand{\ep}{\epsilon}
\newcommand{\pa}{\partial}

\newcommand{\lp}{\left(}
\newcommand{\rp}{\right)}

\newcommand{\dtr}{\dd^{3}{\rm r}}

\newcommand{\maw}{\mathcal{W}}
\newcommand{\mak}{\mathcal{K}}

\newcommand{\mai}{\mathcal{I}}

\newcommand{\mapp}{\mathcal{P}}
\newcommand{\maa}{\mathcal{A}}

\documentclass[10pt,fleqn]{article}
\usepackage[breaklinks]{hyperref}
\usepackage[german,british]{babel}
\usepackage{fancyheadings}
\usepackage{amssymb,amsmath,amsthm}
\usepackage{latexsym,picinpar,amscd,graphpap,pstcol,pst-all}
\usepackage{graphicx}
\DeclareFixedFont{\tr}{OT1}{pnc}{}{}{17pt}
\DeclareFixedFont{\trdos}{OT1}{pnc}{}{}{10pt} 
\begin{document}
\title{\Large {\tr Dark Energy in Astrophysical Context}}
\author{{\Large A. Balaguera-Antol\'{\i}nez}\footnote{E-mail: a-balagu@uniandes.edu.co},
{\Large M. Nowakowski}\footnote{E-mail: mnowakos@uniandes.edu.co}\\
{\small \textit{Departamento de F\'{\i}sica, Universidad de los Andes,}}\\ 
{\small \textit{A.A. 4976, Bogot\'a, D.C., Colombia.}}}
\maketitle


\begin{abstract}
We explore local consequences of a non-zero cosmological constant on astrophysical structures. 
We find that the effects are not only sensitive to the density of the configurations  
but also to the geometry. For non-homogeneous configurations, we calculate the effects 
for a polytropic configurations and the isothermal sphere. 
Special emphasis is put on the fact that the cosmological constant sets certain scales of length, time, mass and density. 
Sizable effects are established for non spherical systems such as elliptical galaxy clusters 
where the effects of $\Lambda$ are growing with the flatness of the system. 
The equilibrium of rotating ellipsoids is modified and the cosmological constant allows new configurations of equilibrium. 
\end{abstract}



\section{Introduction: cosmological motivations for Dark Energy}
Astronomical data based on light curves of distant Ia Supernova \cite{sn}, anisotropies in the cosmic background radiation 
\cite{wmap} and the matter power spectrum of large scale structure \cite{sloan} agrees with an accelerated universe 
which is spatially flat and dominated by a Dark Energy component with a contribution of $70\%$. 
The remaining $30\%$ corresponds to a cold dark matter component, while the contribution of radiation 
and baryons is negligible. In order to account for an accelerated phase at the present epoch, 
the dark energy component must have a negative equation of state $p_{\rm x}=\omega_{\rm x}\rho_{\rm x}$, $\omega_{\rm x}<0$. 
The most favored scenario is the model with a cosmological constant 
$\Lambda$ giving $p_{\rm x}=p_{\rm vac}=-\rvac=\Lambda/8\pi$ (we set the
Newtonian constant, $G_N$, to one), 
called $\Lambda$CDM model. However there are other interesting models introduced in order to mimic the 
current contribution of the Dark Energy component. 
Some of the most relevant models are i) Dark Energy with $p_{\rm x}=\omega_{\rm x}\rho_{\rm x}$ where $-1<\omega_{\rm x}<0$, 
ii) Dynamical Dark Energy (DDE) where the equation of state is a function of time $\omega_{x}=\omega_{\rm x}(t)$, 
iii) Chaplygin Gas \cite{chap} with an equation of state of the form 
$p=-\alpha H_{0}^{2(\gamma +1)}\rho^{-\gamma}$ where $\alpha < (3/8\pi)^{\gamma +1}$, a consequence of $\ddot{a} >0$ and
finite age of the universe,
iv) higher order gravity \cite{hog} and v) Quintessence models \cite{qss} 
which introduce the concept of scalar fields ruled by a potential $V(\phi)$ 
which at the present time mimics a cosmological constant 
as $\Lambda \sim V(\phi_{0})$ where $\phi_{0}$ is the value of the field at the minimum of the potential. 
Although in principle one could translate one model into the other, it is worth noting that 
Quintessence models have been not only successful when trying to interpret the cosmological constant but also in explaining 
the well known problems of standard cosmology (i.e, the horizon problem, the homogeneity problem and fine-tunning problems) 
\cite{lyth}.  \\
\noindent
After the discovery of the current state of the universe ( the fact that we are dominated by the cosmological constant), 
several question have arisen. The inferred values for cosmological parameters 
such as the density parameter associated with the cold dark matter $\Omega_{\rm cdm}=0.27\pm0.04$ 
and the dark energy component associated with e.g.the  cosmological constant $\Omega_{\rm vac}=0.73\pm 0.04$ \cite{sn} lead 
to the coincidence problem. 
One such coincidence emerges if we compare the scales of length and time for the universe with 
values set by the cosmological constant. 
The length (time) scale set by $\Lambda$ is defined as $r_{\Lambda}\equiv \Lambda^{-1/2}$
which corresponds to $4\times 10^{3} {\rm Mpc}$ ($9.6 {\rm Gyr}$). These scales coincide 
with the \emph{size} (age) of the universe which is proportional to the Hubble's 
distance (time) $d_{H}=H_{0}^{-1}\approx 6\times 10^{3}{\rm Mpc}$ ($14{\rm Gyr}$). 
Essentially the coincidence is that we are living in an epoch dominated by $\Lambda$ where e.g. 
$r_{\Lambda} \sim H_0^{-1}$, a relation possible
only at certain times of the expansion. Other aspects of some astrophysical scales set by $\Lambda$ can be found in
\cite{we1}.
Nowadays the origin of the cosmological constant and related problems are a very active field of modern physics \cite{papers}.

\noindent In standard cosmology the universe is considered  an isotropic and homogeneous fluid at large scales. 
This allows us to describe it by the Robertson-Friedmann-Walker metric 
$\dd s^{2}=-\dd t^{2}+a(t)^{2}\lp(1-kr^{2})^{-1}\dd r^{2}+r^{2}\dd \Omega^{2}\rp$ 
where $k$ represents the curvature, $\dd \Omega^{2}$ is the metric on a $2$-sphere and $a(t)$ is the dimensionless 
scale factor. 
This factor determines the kinematics of the universe though the Einstein field equations. 
For a spatially flat universe one obtains the Friedmann equation 
$(\dot{a}/a)^{2}\equiv H^{2}(z)=H_{0}^{2}\Omega_{\rm vac}h_{1}(z)$ 
and acceleration equation $\ddot{a}/a=H_{0}^{2}\Omega_{\rm vac}h_{2}(z)$. 
The time variable $z$ is the red-shift defined as $z=a^{-1}-1$. 
The functions $h_{1,2}(z)$ are given as 
\begin{equation}\label{fri}
h_{1}(z)\equiv \frac{\Omega_{\rm cdm}}{\Omega_{\rm vac}}(1+z)^{3}+(1+z)^{f(z)},
\hspace{1cm}h_{2}(z)\equiv -\frac{1}{2}
\left[\frac{\Omega_{\rm cdm}}{\Omega_{\rm vac}}(1+z)^{3}+(1+z)^{f(z)}\lp1+3\omega_{\rm x}(z)\rp\right],
\end{equation}
We have only considered contributions from a cold dark matter component and vacuum energy density. 
The function $f(z)$ is given by 
\begin{equation}
f(z) = -\frac{3}{\ln (1+z)}\int_{0}^{z}\frac{1+\omega_{\rm x}(z')}{1+z'}\dd z', 
\end{equation}
which follows from the energy-density conservation law 
$\dot{\rho}=-3H(t)(\rho+p)$ for the dark energy component $p_{\rm x}(z)=\omega_{\rm x}(z)\rho_{\rm x}(z)$. 
As pointed before, the case $\omega_{\rm x}=-1$ 
corresponds to the cosmological constant $\rho_{\rm x}=\rvac$. Alternative models with 
$\omega_{\rm x}<-1$ lead to future singularities in the so-called Phantom regime \cite{phantom}.\\

\section{Local Dynamics with Dark Energy}
Local effects of Dark Energy have been investigated by many authors \cite{local}.
We explore these effects of the background through the Newtonian limit of Einstein field 
equations \cite{marek1} from which one can derive a modified Poisson's equation 
$\nabla^{2}\Phi=4\pi \delta \rho-3H_{0}\Omega_{\rm vac}h_{2}(z)$, 
where $\delta\rho$ is the overdensity that gives rise to the gravitational source of the potential $\Phi$. 
The total energy density within the clustered configuration is a contribution of the cold dark matter 
in the background and the collapsed fraction, i.e,  
$\rho(z)=\rho_{\rm cdm}(0)(1+z)^{3}+\delta \rho$.  The extra-term $3H_{0}\Omega_{\rm vac}h_{2}(z)$ in Poisson's equation 
reduces to $8\pi \rvac$ for $\omega_{\rm x}=-1$ if we neglect the contribution from the cold dark matter component. This is 
the standard Newtonian limit. The more general limit retaining also the time ($z$) dependent parts is derived e.g. in
\cite{adotdot}.

\noindent The solution for the potential can be simplified as 
$\Phi(r,z)=\Phi_{\rm grav}(r)-H_{0}^{2}\Omega_{\rm vac}h_{2}(z)r^{2}$
which defines the Newton-Hooke spacetime \cite{nh} for $\omega_{\rm x}=-1$ and $\rho_{\rm cdm}\sim 0$ at the present time. 
Two effects combine in order that we are allowed to make this approximation at the present time: 
i)the red-shift-dependence $(1+z)^{3}$ and the ratio $\rho_{\rm cdm}/\rho$ with $\rho$ 
the mean density of the configuration. This ratio was $10^{-2}$ at the moment of virialisation.\\
Once one knows the solution of Poisson equation $\Phi(r,z)$, one can write the Euler's equation for the overdensity 
as $\delta \rho \dot{u}_{i}+\partial_{j}\mapp_{ij}+\delta \rho \partial_{i} \Phi=0$, 
(neglecting $\rho_{\rm cdm}$), where $\mapp_{ij}$ is the dispersion tensor. 
Taking spatial moments on Euler's equation one can derive the second order tensor virial equation \cite{we3} as
\begin{eqnarray}
\label{tvir}
\frac{1}{2}\frac{\dd ^{2}\mai_{ik}}{\dd t^{2}}&=&\frac{1}{2}H_{0}^{2}\Omega_{\rm vac}(1+z)\left[
h_{1}(z)(1+z)\frac{\dd ^{2}\mai_{ik} }{\dd z^{2}}+\lp h_{2}(z)+2h_{1}(z)\rp\frac{\dd \mai_{ik} }{\dd z}
\right] \nonumber \\
&=&2T_{ik}-|\maw_{ik}^{\rm grav}|
+H_{0}^{2}\Omega_{\rm vac} h_{2}(z)\mai_{ik}+ \Pi_{ik},
\end{eqnarray} 
where we have neglected surface integrals (this is not the case when we consider configurations with a non zero pressure at 
the boundary or if we consider non radial external forces, as electromagnetic forces). 
The first line of (\ref{tvir}) is the second derivative with respect to time of
the inertial tensor in dependence of the red-shift where we have used the Friedmann equation ($h_{1}(z)$)
and the acceleration equation ($h_{2}(z)$). 
The second line of (\ref{tvir}) is the dynamical information encoded in the virial equation. 
The quantities involved in this expression are defined as 
\footnote{We have renamed the density of the configurations from $\delta \rho$ to $\rho$.}
\begin{equation}
\label{ichapter}
\mai_{ik}\equiv \int_{V}\rho r_{i}r_{k}\, \dtr,\hspace{0.5cm}T_{ik}\equiv \frac{1}{2}\int_{V}\rho u_{i}u_{k}\,
\dtr\hspace{0.5cm}\maw_{ik}^{\rm grav}\equiv -
\int_{V}\rho r_{i}\pa_{j}\Phi_{\rm grav} \dtr,\hspace{0.5cm}\Pi_{ik}\equiv \int_{V}\mapp_{ik} \dtr,
\end{equation}
corresponds to the moment of inertia tensor, the kinetic energy tensor, 
the gravitational potential energy tensor and the dispersion tensor. \\
\noindent By taking the trace on Eq. (\ref{tvir}) and assuming the virial graviational equilibrium we 
obtain the most known form for the virial theorem:
\begin{equation}
\label{virial}
|\maw^{\rm grav}|=2T+3\Pi+\frac{8}{3}\pi \rvac \mai,
\end{equation} 
where $T={\rm Tr}(T_{ik}$) and so on for the other terms. Note then that the effects of a positive 
cosmological constant are \emph{coupled} to the moment of inertia 
which depends on the density  profile (or the equation of state) 
and the geometry of the configuration. Hence, these two parameters (density and geometry) become relevant 
in order to determine sizable effects of a vacuum energy density. 
It is to be expected that some effects due to $\Lambda$ will show up only if the density of the 
object is low. However, we will see that geometry helps to enhance such effects.\\
\noindent In order to describe a general consequence of Eq.(\ref{virial}), we write the total kinetic energy as $\mak=T+3\Pi$. 
Since this quantity is positive defined, we conclude that the state of equilibrium is characterized by the 
inequality $\tilde{\varrho}>\maa \rvac$ where $\tilde{\varrho}$ is a characteristic 
parameter with units of density (such as the man density or the central density) 
and $\maa \equiv (16\pi/3)\tilde{\mai}/|\tilde{\maw}^{\rm grav}|$, 
together with $\tilde{\mai}=\mai/\tilde{\varrho}$ and $|\tilde{\maw}^{\rm grav}|=2|\maw|/\tilde{\varrho}^{2}$. 
The parameter $\maa$ depends on the geometry of the configurations 
and/or the parameters specifying internal structure such as the equation of state.
\section{Effects on astrophysical configurations}
\subsection{Spherical configurations}

The tensor virial equation is trivially satisfied for spherically symmetric configuration 
(in the absence of electromagnetic fields or non radial forces)
since $\maw_{ik}^{\rm grav}=\delta_{ik}\maw^{\rm grav}$ and $\mai_{ik}=\delta_{ik}\mai$. 
Therefore Eq.(\ref{virial}) is sufficient in order describe configurations with this geometry.
\subsubsection{Homogeneous systems} 
For constant density (with $\tilde{\varrho}=\rho$) it is straightforward to check that $\maa_{\rm sph}=2$. 
In this case the effects of $\rvac$ are not enhanced because of this geometrical factor. 
However, other effects can be shown to exist. 
Indeed, the presence of a cosmological constant allows the existence of a maximal virial radius of a
spherical configuration in virial equilibrium in the limit when $\mak\to 0$. The radius at virial equilibrium can be
calculated  from Eq. (\ref{virial}), 
as the solution of the cubic equation $R_{\rm vir}^{3}+\lp10\eta \rl^{2}\rp R_{\rm vir}-3M \rl^{2}=0$ 
where $\eta\equiv \mak/M=(3k_{B}/\mu)T$, $\mu$ is the mass of the average member of the configuration, 
$k_{B}$ is the Boltzmann constant, $T$ is the temperature. 
The solution for the radius at virial equilibrium reads as  \cite{we3}
\begin{equation}
\label{delta}
R_{\rm vir}(x,\eta)=2.53 x^{-1/3}\eta^{1/2}\sinh\left[\frac{1}{3}{\rm arcsinh}\lp0.24 x\eta^{-3/2}\rp\right]R_{\rm max},
\end{equation} 
where $x=M\rl^{-1}$ and $R_{\rm max}=R_{\rm vir}(x,0)=(3M\rl^{2})^{1/3}$ 
is the maximum radius allowed for astrophysical/cosmological structures. In astrophysical terms we have
\begin{equation}
\label{sca3}
R_{\rm max}=9.5\times 10^{-2}\lp\frac{M}{M_{\odot}}\rp^{\frac{1}{3}}\Omega_{\rm vac}^{-\frac{1}{3}} h_{70}^{-\frac{2}{3}} 
\, {\rm kpc}.
\end{equation}
This length scale is a combination of a cosmological length scale defined by 
the cosmological constant $r_{\Lambda}$ and a pure astrophysical scale, i.e, 
the Schwarszchild's radius $R_{\rm s}= 2M$. This combination gives rise to a relevant astrophysical scale. 
This new scale with a replacement $M \to \mu$ in 
(\ref{sca3}) also appears when studying the motion of test particles in the Einstein-de Sitter 
space time \cite{we2} and represents the last bound orbit around the source of gravity with mass $\mu$. 
In other words, Eq. (\ref{sca3}) would imply that the cosmological constant sets the typical scale for bound 
configurations. Some examples are welcome: for $\mu=10^{6} M_{\odot}$ i.e. a typical mass of a
globular clusters (the tight connection between galaxies and globular clusters is that the latter are 
important in galaxy formation) 
we get $R_{\rm max}\sim 8$ kpc, whereas for a galaxy mass, $\mu=10^{11}M_{\odot}$, 
we obtain  $R_{\rm max}\sim 3\times 10^{2}$ kpc. 
The numerical outcomes correspond to the radius a typical galaxy in the first example, 
while the second one agrees with numerical values for the extension of a galaxy cluster.
\subsubsection{Non-homogeneous systems}
\begin{figure}[t]
\begin{center}
\includegraphics[angle=270,width=12cm]{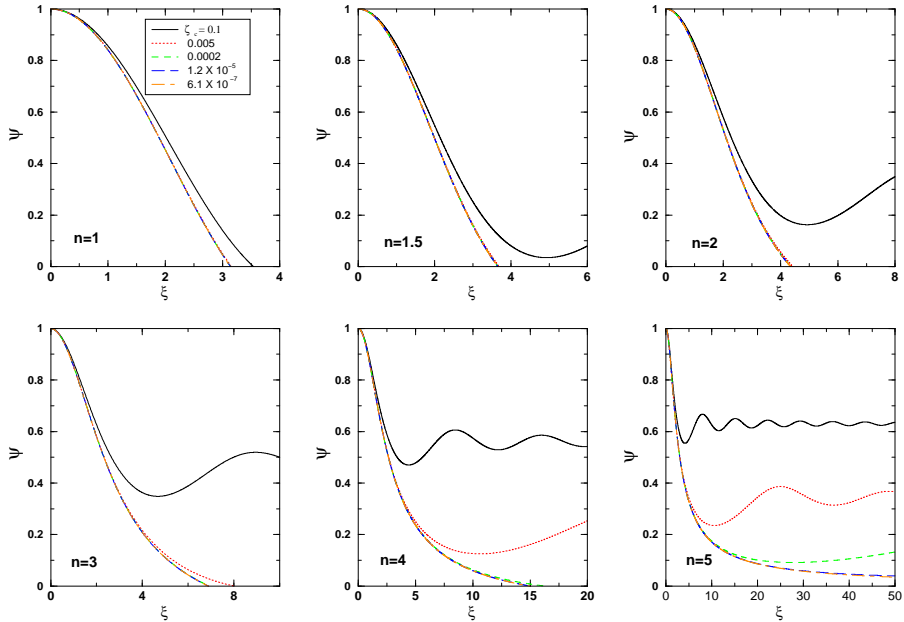}
\caption{\footnotesize{Solutions of Lane-Emden equation for different 
indexes $n$ and ratios $\zeta_{\rm c}=2\rvac/\rho_{\rm c}$ \cite{we3}.}}
\label{lane}
\end{center}
\end{figure}

The effects on non-homogeneous spherical configurations have been explored by considering 
polytropic configurations characterized by a barotropic equation of state of the form 
$p=\kappa \rho^{1+1/n}$ \cite{chan}. 
The equation governing the behavior of the density of polytropic 
configurations with cosmological constant is the modified Lane-Emden equation written as
\begin{equation}
\label{le}
\frac{1}{\xi^{2}}\frac{\dd }{\dd \xi}\lp \xi^{2}\frac{\dd \psi}{\dd \xi}\rp=\zeta_{\rm c}-\psi^{n},
\end{equation} 
where $\xi=r/a$, $\psi=(\rho/\rho_{\rm c})^{1/n}$, $\rho_{\rm c}=\rho(r=0)$, 
$a\equiv (\kappa (n+1)/4\pi)^{1/2}\rho_{\rm c}^{1/2n-1/2}$ and $\zeta_{\rm c} \equiv 2 \rvac/\rho_{\rm c}$.
As a main effect of a positive $\Lambda$ one can demonstrate that not all the configurations 
with a polytropic equation of state have a defined radius, 
even for $n<5$, as can be seen in Fig.\ref{lane}(i.e. not all values of $n$ leads to a solution $\psi$ 
such that for some $\xi_{1}$ one gets $\psi(\xi_{1}=0)$). We generalize 
the equilibrium criteria given for homogeneous configurations by identifying $\tilde{\varrho}=\rho_{\rm c}$ 
and writing $\rho_{\rm c}>\maa_{n}\rvac$ 
where the factor $\maa_{n}$ takes different values for different indexes $n$. 
Some examples are $\maa_{1}=10.24$, $\maa_{3/2}=24.24$, $\maa_{3}=307.69$, $\maa_{4}\approx 4000$. \cite{we5}. 
Characteristic quantities and concepts such as the radius and the mass as well as 
the stability criteria get modified in the presence of $\Lambda$. 
For instance, the radius of such configurations can be written as $R_{n}=\alpha_{n}R_{0}$
where $R_{0}$ is the radius that one would obtain with $\rvac=0$ and $\alpha_{n}=\alpha_{n}(\zeta_{\rm c})$ 
is an enhancement factor which depends on the polytropic index $n$ and the ratio $\zeta_{\rm c}$. 
Some numerical values are $\alpha_{1}(0.1)=1.12$, $\alpha_{3/2}(0.05)=1.11$, $\alpha_{1}(0.001)=1.001$, 
$\alpha_{3}(0.001)=1.022$
which shows that even for spherical configurations effects are found 
for low density configurations with different values of $n$.  Worth noting is the case of the isothermal sphere i.e.
the case $n \to \infty$. The isothermal sphere is an often employed model in astrophysics \cite{binney}. From Figure 1
we can see that including $\Lambda$ the cases $n \ge 5$ do not have a defined radius even in the asymptotic sense.
Therefore one can easily suspect that the extreme case of the isothermal sphere will display even bigger effects which 
is to say, sizable effects for larger density. This is indeed the case and the isothermal sphere is not a viable model
(with $\Lambda$) even at density comparable with galaxy densities \cite{we5}.
The stability criteria for  polytropic configurations can be derived via the virial theorem in two ways: 
one option is to develop Lagrangian perturbations of the form $\xib e^{i\omega t}$ 
(where $\xib$ is the Lagrangian displacement) on the tensor virial equation and derive 
an equation for the oscillation frequency around equilibrium $\omega$, 
such that stability would implies $\omega^{2}>0$ (see for instance \cite{chan2}). 
Another option is to solve for the mass of the configuration as a function of the central density. 
Stability (instability) stands for $\pa M/\pa \rho_{\rm c}>0$ ($\pa M/\pa \rho_{\rm c}<0$) \cite{tas}. 
Both ways lead us to write a critical value for the polytropic 
index such that polytropic configuration are stable if $\gamma>\gamma_{\rm crit}$ where
\begin{equation} 
\label{slope2}
\gamma_{\rm crit}=\gamma_{\rm crit}(\zeta_{c})\equiv \frac{4}{3}+\frac{2}{3}\frac{\pa \ln \mathcal{G}}{\pa \ln \rho_{\rm c}},
\end{equation}
is the critical polytropic index \cite{we5}. 
The function $\mathcal{G}$ depends on the ratio $\zeta_{\rm c}$ through 
integrals of the Lane-Emden function \cite{we5} which in turn depends on the index $n$. 
Hence, Eq (\ref{slope2}) is a transcendental equation for the critical value of the polytropic index. 
Nevertheless, Eq.(\ref{slope2}) is a generalization of stability criteria for 
homogeneous configurations under radial adiabatic perturbations with cosmological constant and can be 
written as $\gamma>(4/3)(1+(1/2)\zeta )$. It is worth mentioning that by including the corrections due 
to general relativity, the critical value for $\gamma_{\rm crit}$ is also 
modified as $\gamma_{\rm crit}=(4/3)+R_{\rm s}/R$ \cite{shapiro} and hence for compact objects the 
correction to the critical polytropic index is stronger from the effects of general relativity than 
from the effects of the background, as expected. 

\subsection{Non-spherical configurations}
\begin{figure}
\begin{center}
\label{ag}
\includegraphics[angle=270,width=10cm]{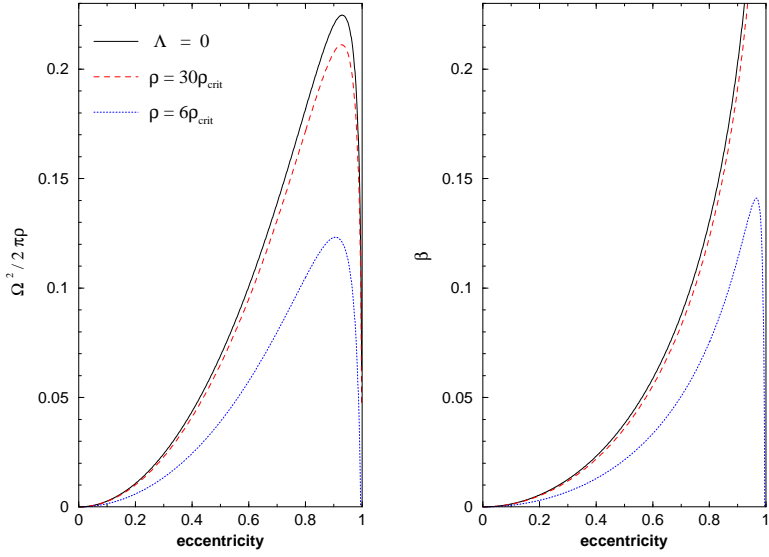}
\caption{\footnotesize{Angular velocity and the function $\beta$ as defined in the text versus
eccentricity $e$.}}
\end{center}
\end{figure}

As mentioned before, the effects of a positive cosmological constant can be enhanced through the 
geometrical factor $\maa$. As an example, we consider the virial theorem and solve for the velocity dispersion for arbitrary
geometry and constant density. The final expression has the form $v^{2}/v^{2}_{\Lambda=0}=1-(1/2)\zeta \maa$ where
$\zeta=2\rvac/\rho$. In the last section we saw that for spherical and homogeneous 
systems $\maa_{\rm sph}=2$ and hence the correction is just $1-\zeta$ which implies that a sizable 
effect could be reached only in systems with $\rho\to 2\rvac$. 
On the other hand, for non-spherical configurations we can combine a 
low density feature with a highly non-spherical geometry in order to find sizable effects on realistic systems. For
instance, for oblate and prolate homogeneous configurations these
geometrical factor is written as \cite{we3,binney}
\begin{eqnarray} \label{A}
\maa_{\rm obl}&=&\frac{4}{3}\lp\frac{3-e^{2}}{\arcsin e}\rp\frac{e}{2\sqrt{1-e^{2}}} 
\stackrel{ a_{1} \gg a_{3}}{\to}  \frac{8}{3\pi}\lp\frac{a_{1}}{a_{3}}\rp, \nonumber \\
\maa_{\rm pro}&=&
\frac{4}{3}\frac{e(3-2e^{2})}{(1-e^{2})^{3/2}}\left[\ln\lp\frac{1+e}{1-e}\rp\right]^{-1} 
 \stackrel{ a_{3} \gg a_{1}    }{\to} \frac{2}{3}\lp\frac{a_{3}}{a_{1}}\rp^{3}\left[\ln\lp\frac{2a_{3}}{a_{1}}\rp\right]^{-1},
\end{eqnarray}
where $e^{2}=1-a_{3}^{2}/a_{1}^{2}$ is the eccentricity 
for oblate configurations and $e^{2}=1-a_{1}^{2}/a_{3}^{2}$ the corresponding one for the prolate case. 
From Eq.(\ref{A}) we see that   very flat astrophysical objects we can gain several orders 
of magnitude of enhancement of the effect of $\rvac$. Clearly we often encounter 
in the universe flat configurations such as elliptical galaxies, 
spiral disk galaxies, clusters of galaxies of different forms and finally 
superclusters which can have the forms of pancakes.
Of course, the more dilute the system is, the bigger the effect of $\Lambda$. 
We can expect sizable effects for clusters and superclusters,
even  for very flat galaxies. 
Regarding the latter, low density galaxies like the nearly invisible galaxies are among other the best candidates.
\\
In order to explore the effects of $\rvac$ further we turn to the tensor virial equation for a 
rotating oblate configuration. Writing $2T_{ik}=\Omega_{\rm rot}^{2}\mai_{ik} -\Omega_{{\rm rot} i}
\mai_{kj}\Omega_{j {\rm rot}}$ as (twice) the rotational kinetic energy tensor 
and defining a coordinate system such that $\vec{\Omega} =\Omega \hat{e}_{z}$ 
one gets the following expression for the angular velocity:
\begin{equation}
\label{ns01}
\Omega_{\rm rot}^{2}=\lp\frac{|\maw_{xx}^{\rm grav}|-|\maw_{zz}^{\rm grav}|}{\mai_{xx}}\rp-
 \frac{8}{3}\pi \rvac\lp1 -\frac{\mai_{zz}}{\mai_{xx}}\rp=\Omega_{0}^{2}\left[1-\frac{1}{2}\zeta g(e)\right],
\end{equation}
where $\Omega_{0}$ corresponding to the angular velocity when $\Lambda=0$ given 
by the Maclaurin formula (see \cite{chan}) and $\zeta \equiv 2 \rvac/\rho$. The function $g(e)$ has been defined as 
\begin{equation}
\label{H}
g(e)\equiv\frac{4}{3}e^{5}\left[(1-e^{2})^{1/2}(3-2e^{2})\arcsin e-3e(1-e^{2})\right]^{-1}.
\end{equation}
As can be seen from these expressions, the vacuum energy density $\rvac$ has a two effects on the angular velocity.
In first case, it reduces the angular velocity with respect to the value $\Omega_{0}$ especially 
at the local maximum (see Figure \ref{ag}). This is not a small effect and can affect even galaxies. 
Secondly, we see from Eq. (\ref{H}) that in the limit $e \to 1$ 
one has $(1/2)\zeta g(e) \to (1/2)\zeta (32\pi/9)\left(a_{3}/a_{1}\right)$, 
approaching $1$ for a very flat oblate configuration and not too dense matter. 
Therefore, beyond the local maximum in $\Omega_{\rm rot}$ the cosmological constant causes a  
steeper fall of $\Omega_{\rm rot}$ towards $0$. In passing we note that galaxy clusters exhibit rotating ellipsoidal 
configurations \cite{clusters} to which our results can be applied.\\
Figure 2 shows the angular velocity and the ratio $\beta$
defined as the ratio of rotational over gravitational contributions i.e. $\beta=T/|\maw^{\rm grav}|$
versus the eccentricity $e$.\\
The tensor virial equation not only gives us the equation for angular velocity 
(which can be also derived from Euler's equation) but also constrains 
the non-spherical geometries in equilibrium. 
That is, not all values of eccentricities represent configurations with rotation which are 
in equilibrium \cite{chan2}. For instance, with $\rvac=0$ only two configurations are 
allowed to have angular velocity along the minor axis and to be in equilibrium, namely, 
the Maclaurin solution (oblate) with $0<q_{3}<1$ and the Jacobi solution (triaxial) 
with $0<q_{3}<0.582$ where $q_{3}=a_{3}/a_{1}$ (the value $q_{3}^{\rm bif}=0.582$ is known as the bifurcation point). 
With the inclusion of $\rvac$ in the virial equation, we find two effects \cite{we4}. 
In first place, we see that the bifurcation point becomes a increasing function of $\zeta$. 
Secondly, we find a second bifurcation point $q_{3(\rm min)}^{\rm bif}$ 
such that for $q_{3}<q_{3(\rm min)}^{\rm bif}$, the only allowed solution is again the Maclaurin solution.\\
Similar results are found for triaxial configurations which can be reduced 
to prolate ellipsoids and angular velocity in the direction of the largest axis which is called minor axis
rotation. This is rare case, but can nevertheless be found in nature \cite{minorot}.

\section{Rotating configurations in an expanding background}
In this section we consider the full contribution of the background. 
This is to say we examine how virialized matter responses to a time dependent external force representing the expansion.
We assume that the shape of the system remains constant even under the effects of a time dependent background. 
It is clear that a time dependent term in the virial equation implies that certain internal properties of
the matter distribution (like angular velocity and internal velocity) also change in time, in such a way as to maintain
a constant volume and density. 
We concentrate here on the effects on angular velocity. Using equation (1) and (3) 
and assuming the left hand side of (3) to be zero
(i.e. we assume gravitational equilibrium) we obtain
\begin{equation}\label{ange}
\ep=\ep(z;\omega_{\rm x},\zeta,e)\equiv \frac{\Omega^{2}-\Omega^{2}_{0}}{\Omega_{0}^{2}}=\frac{1}{4}g(e)
\zeta\left[\frac{\Omega_{\rm cdm}}{\Omega_{\rm vac}}(1+z)^{3}+\lp 1+3\omega_{\rm x}(z)\rp (1+z)^{f(z)} \right].
\end{equation}
where $\Omega_0$ is the angular velocity calculated without any Dark Energy background.
As mentioned above in order to maintain a constant shape in a expanding background the system has to pay the price 
that its angular velocity will change with time, according to the dominant component of the background. 

\begin{figure}
\begin{center}
\label{g1}
\includegraphics[angle=270,width=10cm]{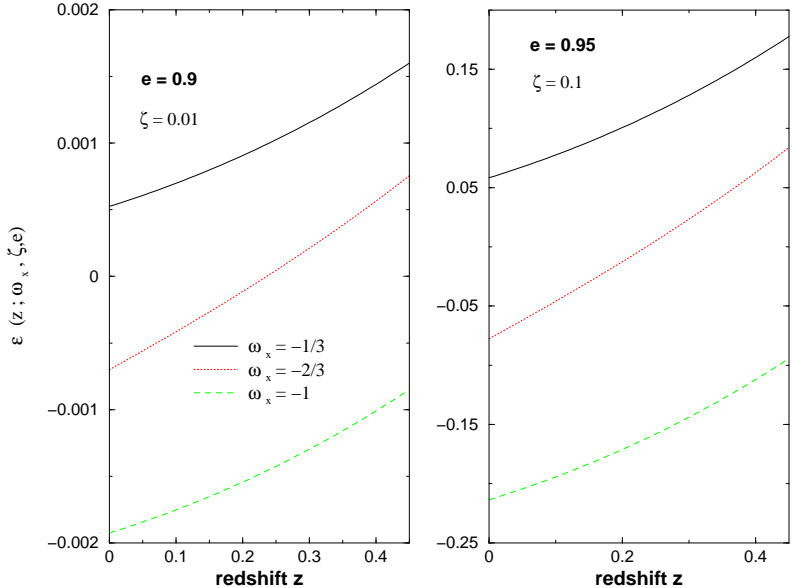}
\caption{\footnotesize{The fractional variation of the angular velocity with respect its 
value for $\Lambda=0$ as a function 
of the redshift for different Dark Energy models with typical values of $\zeta$ and eccentricities. }}
\end{center}
\end{figure}

If we neglect the CDM component with respect to the density of the system, we see that the Dark Energy component 
makes the angular velocity to decrease in order for the system to maintain equilibrium 
since $1+3\omega_{\rm x}<0$. At $z=0$ we get $\ep\sim -0.4g(e)\zeta$. 
An extreme situation would be reached if $\ep\to 1$ so that $\Omega^{2}\to 0$. 
For $e\sim 0.8$ this would require $\zeta\sim 0.5$, which is a very diluted configuration. 
For higher eccentricities (say $e\sim 0.97$) one would need a system with $\zeta\sim 0.2$ which is still a rather low value. 
In the $\Lambda$CDM model $\Omega=0$ is reached at a redshift given by 
$z_{c}=\left[(2\Omega_{\rm vac}(\zeta g(e)-2))/(\zeta g(e)\Omega_{\rm cdm})\right]^{1/3}-1$. 
For realistic examples as  $\zeta\sim 0.023$ (this represents the case of a configurations with $\rho\sim 200 \rho_{\rm cdm}$
which is the typical value predicted by the nonlinear collapse model) an allowed redshift could be reached if
$e$ is very close to $1$. This clearly means the extreme situation of $z_{c}$ requires extreme flat objects. 
In Fig. \ref{g1} we show the behavior of $\ep$ as a function of the redshift with $\zeta=0.01,0.1$ and $e=0.9,0.95$ 
for three different values of the equation of state for Dark Energy. 
It is clear that the effects associated with a cosmological constant are 
stronger than the ones associated with other Dark Energy models, 
but in general those effects are small for realistic values of $\zeta$ as 
the used in the figure. We would have to measure the angular 
velocity at different red-shifts very exactly to see an effect over a range of $z$.
However, the difference between the models is more significant.  

\section{Conclusions} 
In the present work we investigated in detail the astrophysical relevance of the 
cosmological constant for equilibrium configurations. Using the tensor and scalar virial equations and the
Lane-Emden equation we could show that many astrophysical facets get modified by $\Lambda$.
An important aspect is the fact that $\Lambda$ introduces new relevant scales (these scales
would be zero if $\Lambda=0$) like the maximal
virial volume defined by the maximal virial radius (\ref{delta}) and the maximal extension of bound orbits.

It is often assumed that superclusters with densities $\sim \rho_{\rm crit}$ 
are not in equilibrium. With the inequality $\tilde{\varrho}>\maa \rvac$ we have a precise tool to quantify this statement. 
Indeed, the pancake structure of the superclusters lead us to the conclusion 
that they are even far away from the equilibrium state due to the factor $\maa$ 
which grows with the object's flatness. On the other hand we can apply the results also
to low density clusters which are assumed to be virialized. Depending on shape and density
various observable, like internal velocity or angular velocity, will get affected by the
repulsive force induced by Dark Energy. The effects are bigger the large the geometry of the
objects deviates from spherical symmetry and the lower the density of the object. Finally, the response of the
objects to epoch dependent expansion gets reflected also in a time dependent variation of some of their
properties and the difference between different Dark Energy models becomes manifest.

\section*{Acknowledgments}
We thank our collaborators, C. B\"ohmer and D. F. Mota, for contributions and numerous discussions.

\end{document}